\documentclass[12pt]{article}
\usepackage{amsmath,amsfonts,amssymb}
\usepackage{hyperref}
\usepackage{color}
 
\def\bS{\mathbf{S}}
\allowdisplaybreaks
\usepackage{graphicx}
\newcommand{\beq}{\begin{equation}}
\newcommand{\eeq}{\end{equation}}
\newcommand{\beqa}{\begin{eqnarray}}
\newcommand{\eeqa}{\end{eqnarray}}
\newcommand{\beqar}{\begin{eqnarray*}}
\newcommand{\eeqar}{\end{eqnarray*}}

\newcommand{\eps}{\epsilon}

\newcommand{\nn}{\nonumber}

\newcommand{\z}{\zeta}

\newcommand{\ie}{{\it i.e.,}\ }
\newcommand{\labell}[1]{\label{#1}} 
\newcommand{\reef}[1]{(\ref{#1})}

\newcommand\Tr{{\rm Tr}}
\newcommand\tr{{\rm tr}}

\usepackage{amsmath,amsfonts,amssymb}
\def\nn{\nonumber}
\def\bS{\boldsymbol{S}}

\begin{document}
\begin{titlepage}
\begin{center}

\vskip 2 cm
{\LARGE \bf $R^4 $ terms in   supergravities   \\  \vskip 0.25 cm  via T-duality constraint
 }\\
\vskip 1.25 cm
 Hamid Razaghian\footnote{razaghian.hamid@gmail.com}  and  Mohammad R. Garousi\footnote{garousi@um.ac.ir}

\vskip 1 cm
{{\it Department of Physics, Faculty of Science, Ferdowsi University of Mashhad\\}{\it P.O. Box 1436, Mashhad, Iran}\\}
\vskip .1 cm
 \end{center}

\begin{abstract}
 It has been   speculated in the literature that the   effective actions of  string theories  at any order of $\alpha'$ should be invariant under the Buscher rules plus their higher covariant derivative corrections.   This  may be used as a constraint to find effective actions  at any order of $\alpha'$, in particular,    the metric, the B-field and the dilaton  couplings in   supergravities at order $\alpha'^3$ up to an overall factor. For the simple case of zero B-field and diagonal metric in which we have done the calculations explicitly, we have found that  the constraint fixes almost all the seven independent Riemann curvature couplings. There is only one term which is not fixed, because    when metric is diagonal, the reduction of  two $R^4$ terms   become identical.   The Riemann curvature couplings that the T-duality constraint produces for both  type II   and    Heterotic theories are fully consistent with the existing couplings in the literature which have been found by the S-matrix   and by the sigma-model approaches.
\end{abstract}
\end{titlepage}

\section{Introduction}

String theory is a candidate for the quantum gravity which produces the classical supergravity at low energy. The stringy  signature of the quantum gravity appears in the higher derivative corrections to the supergravity.   There are various   techniques  in string theory for extracting these higher derivative corrections. Scattering amplitude   approach \cite{Scherk:1974mc,Yoneya:1974jg}, sigma-model approach \cite{Callan:1985ia,Fradkin:1984pq,Fradkin:1985fq},  supersymmetry approach \cite{Gates:1986dm,Gates:1985wh,Bergshoeff:1986wc,Bergshoeff:1989de}, Double Field Theory  (DFT) approach  \cite{Siegel:1993xq,Hull:2009mi,Hohm:2010jy},  and duality approach \cite{Ferrara:1989bc,Font:1990gx,Garousi:2017fbe}. In the duality approach, the  consistency of the effective actions with duality transformations are imposed to find the higher derivative couplings \cite{Garousi:2017fbe}. In particular, it has been speculated that the consistency of the effective actions at any order of $\alpha'$ with the T-duality transformations may  fix both the effective actions and the T-duality transformations   \cite{Razaghian:2017okr}. 

The T-duality in string theory is realized by studying the spectrum of the closed string on a tours. The spectrum is invariant under the transformation in which the Kaluza-Kelin modes and the winding modes are interchanged, and at the same time the set of scalar fields parametrizing the tours  transforms to another set of scalar fields parametrizing  the dual tours.  The transformations on the scalar fields have been extended to curved spacetime with background fields by Buscher \cite{Buscher:1987sk,Buscher:1987qj}. It has been observed that the effective actions at the leading order of $\alpha'$ are invariant  under the Buscher rules \cite{Sen:1991zi,Bergshoeff:1995as}. The effective actions at the higher order of $\alpha'$ are also expected to be invariant under the T-duality transformations which are the Buscher rules and their appropriate $\alpha'$  corrections.  These corrections at order $\alpha'$ have been found in  \cite{Bergshoeff:1995cg,Kaloper:1997ux}. 


In type II superstring theory, the   higher derivative corrections to the supergravity begin at order $\alpha'^3$. As a result, the   corrections to the Buscher rules also begin at order $\alpha'^3$.  Hence one expects the effective actions of $O$-plane in the type II superstring theory at order $\alpha'^2$ to be invariant under the Buscher rules. This may be used as a constraint to find the $O$-plane effective actions. The NS-NS couplings on the world volume of  the $O$-plane at order $\alpha'^2$ have  been found in \cite{Robbins:2014ara,Garousi:2014oya} by this constraint. 

The T-duality constraint has been used in \cite{Razaghian:2017okr} for the bosonic  theory to find the effective actions at order $\alpha',\alpha'^2$ and their corresponding T-duality transformations when B-field is zero. Even though,  the constraint does not  completely fix the corrections to the Buscher rules, it however fixes the effective actions which are exactly the same as the effective actions that have been found by the S-matrix and sigma-model approaches up to an overall factor. In this paper, we are going to examine the T-duality constraint  at order $\alpha'^3$. The bosonic, the Heterotic and the type II theories all have corrections at order $\alpha'^3$. However, we are interested in the Heterotic and the type II theories in this paper. In the type II theory, we will find that the T-duality constraint almost fixes the effective action up to an overall factor, whereas, there remain may residual T-duality parameters in the T-duality transformations. In the Heterotic theory, we will again find that the constraint almost fixes the effective action while leaving many parameters in the T-duality transformations at orders $\alpha', \alpha'^2,\alpha'^3$. In this case, however, there are gravity couplings which are resulted from the Green-Schwarz mechanism \cite{Green:1984sg}. Constraining these couplings to be invariant under the T-duality transformations may fix the residual parameters in the T-duality transformations. We will find, by explicit calculations, that this constraint fixes the residual parameters   at order $\alpha'$.

The outline of the paper is as follows: In section 2, we explain our strategy for implementing the T-duality constraint on the effective actions, and discuss our speculation that the T-duality constraint at any order of $\alpha'$ may be used only on the specific  Riemann curvature couplings at that order which are invariant under field redefinitions.  In section 3, we write all independent Riemann curvature couplings at order $\alpha'^3$ and show that they all are invariant under the field redefinitions in type II theory, whereas, two of them are not invariant under the field redefinitions in the Heterotic theory. In section 3.1, we impose the T-duality constraint on the couplings in the type II theory and show that even though the T-duality at order $\alpha'^3$ can not fix all parameters in the T-duality transformations, but it can almost fix all the Riemann curvature couplings up to an overall factor. Since the reduction of two 10-dimensional independent Riemann curvature couplings produces identical 9-dimensional couplings, the T-duality constraint can fix the coefficient of the sum of these two terms. For one particular choice for one of these unfixed coefficients, we show that the couplings that the T-duality constraint produces are exactly the same as the couplings that the S-matrix and sigma-model approaches produce, up to an overall factor. 

 In section 3.2, we impose the T-duality constraint on the couplings in the Heterotic theory and show that the constraint almost fix the Riemann curvature couplings that are invariant under the field redefinitions. The T-duality constraint related the coefficient of the two Riemann curvatures couplings that are not invariant under the field redefinitions, to the T-duality invariant couplings at order $\alpha'$. For the particular couplings at order $\alpha'$ which do not change the graviton and dilaton propagators, we find that the couplings that the T-duality constraint produces are the same as the couplings in the literature, up to an overall factor. In this section, we also show that the gravity couplings which are resulted from the Green-Schwarz mechanism  are invariant under the T-duality transformations. We show that this constraint fixes the residual T-duality parameters at order $\alpha'$. 
 
\section{Strategy}

The higher derivative couplings involving graviton  and dilaton in the effective action at order $\alpha'^n$   can be classified as 
\beqa
S_n=S_n^{(1)}+S_n^{(2)}\labell{S12}
\eeqa
where $S_n^{(1)}$ contains the couplings which are unambiguous, and $S_n^{(2)}$ contains the couplings which are  ambiguous as their coefficients are   changed under field redefinitions.  In general, $S_n^{(1)}$ contains  Riemann curvature couplings with some specific contraction of indices, whereas  $S_n^{(2)}$ contains Riemann curvature couplings with some other contraction of indices and contains Ricci and scalar curvatures as well as dilaton.  Using field redefinitions, one can rearrange  the couplings in $S_n^{(2)}$ into two parts. One part contains the couplings which are invariant under the field redefinitions and the second part contains the couplings which are arbitrarily changed under the field redefinitions. These later couplings may or may not  be   zero   depending on the field variables. For example, 
 at order $\alpha'$, it has been shown in  \cite{Metsaev:1987zx}  that    there are  8 ambiguous coefficients and they  satisfy one relation which is invariant under the field redefinitions. So one can set all the ambiguous coefficients to zero  except one of them. So $S_n^{(2)}$in this case can be simplified to have only one coupling.  At order $\alpha'^2$, there are 42 ambiguous coefficients. They satisfy 5 relations which are invariant under the field redefinitions \cite{Bento1989,Bento:1989ir,Bento1990,Bento:1990jc}.  So one can fix all ambiguous coefficients to zero except 5 of them. As a result, $S_n^{(2)}$ in this case can be written in terms of only five couplings.  Similarly for couplings at higher order of $\alpha'$. Therefore, using field redefinitions, one can write $S_n^{(2)}$ as
\beqa
S_n^{(2)}&=&\sum_{i=1}^m\z_i f_i +\cdots\labell{S2}
\eeqa
where   $f_1,f_2,\cdots$ in the first part are the couplings that their coefficients $\z_1,\z_2,\cdots$ are invariant under the field redefinitions. The dots in above equation represents the  second part which contains the couplings which can be set to zero for specific field variables. The coefficients $\z_1,\z_2,\cdots$ may be fixed by S-matrix calculations.

We now show that consistency of the effective actions with T-duality constrains the coefficients  $\z_1,\z_2,\cdots$ to be zero. If one dimensionality reduces the $D$-dimensional effective action $S_n^{(2)}$ to $ d$-dimensional effective action $\bS_n^{(2)}$ where $D=d+1$, the functions  $f_1,f_2,\cdots$ each can have terms with odd number of $\sigma$ where $\sigma=(\ln G_{yy})/2$. We  call them $f_i^{\rm odd}$. There are also terms with even number of $\sigma$ which we call them $f_i^{\rm even}$. Under the Buscher rules, \ie
\beqa
\sigma&\rightarrow &-\sigma \nn\\
P&\rightarrow &P \nn\\
g_{ab}&\rightarrow &g_{ab} \labell{buscher}
\eeqa
where $P$ and $ g_{ab}$ are  the $d$-dimensional dilaton and metric, respectively, $f_i^{\rm even}$ is invariant and $f_i^{\rm odd}$ changes its sign. 
Then the transformation of $\bS_n^{(2)}$ under the Buscher rules becomes
\beqa
\delta \bS_n^{(2)}&=&2\sum_{i=1}^m\z_i f_i^{\rm odd} +\cdots\labell{dS}
\eeqa
Since we have already used the $D$-dimensional field redefinition to write the action $S_2$ as in \reef{S2} in which the coefficients are invariant under the field redefinitions,  the $d$-dimensional field redefinition does not change the coefficients  $\z_1,\z_2,\cdots$. 
Now using the observation that the $d$-dimensional effective action must be invariant under the Buscher rules up to $d$-dimensional field redefinitions, one concludes that 
\beqa
\z_1=\z_2=\cdots=\z_m=0
\eeqa
As a result, the T-duality constraint on the  effective  action fixes the $S_n^{(2)}$ part of the effective action to be zero up to field redefinitions. Explicit calculations at orders $\alpha'$ and $\alpha'^2$ conform the above conclusion \cite{Razaghian:2017okr}.

Now let us   consider the $S_n^{(1)}$ part of the effective action \reef{S12}. Unlike the $S_n^{(2)}$ part, the coefficients of the couplings in $S_n^{(1)}$ are not changed under the $D$-dimensional field redefinitions, however, after reducing them to the $d$-dimension effective action $\bS_n^{(1)}$, the coefficients of the $d$-dimensional couplings are changed under the $d$-dimensional field redefinitions. Under the dimensional reduction, one can write $\bS_n^{(1)}=\bS_n^{(1) \rm odd}+\bS_n^{(1) \rm even}$ where $\bS_n^{(1) \rm odd}$ contains the terms with odd number of $\sigma$ and $\bS_n^{(1) \rm even}$ contains the terms with even number of $\sigma$. Under the Buscher rules, 
\beqa
\bS_n^{(1) \rm even}&\rightarrow &\bS_n^{(1) \rm even}\nn\\
\bS_n^{(1) \rm odd}&\rightarrow &-\bS_n^{(1) \rm odd}
\eeqa
Then the transformation of $\bS_n^{(1)}$ under the Buscher rules becomes
\beqa
\delta \bS_n^{(1)}&=&2\bS_n^{(1) \rm odd}
\eeqa
It must be zero up to the $d$-dimensional field redefinitions. The $d$-dimensional field redefinitions can be interpreted as higher derivative corrections to the Buscher rules.  Since $\delta \bS_n^{(1)}$ contains only the terms with odd number of $\sigma$, the appropriate field redefinition should produce also terms with odd number of $\sigma$.   If one does not use the $d$-dimensional field redefinitions, then one would find that $S_n^{(1)}$ is zero which is not correct. The   $d$-dimensional field redefinitions add some extra terms to the above equation which makes $S_n^{(1)}$ not to be zero. In fact the resulting constraint may fix the effective action $S_n^{(1)}$ up to an overall factor. In the rare cases that some of the $D$-dimensional independent couplings produce identical $d$-dimensional couplings in  $\bS_n^{(1) \rm odd}$, the above constraint can fix only the   sum of their corresponding coefficients.  It has been shown in \cite{Razaghian:2017okr} that  $S_n^{(1)}$ at orders $\alpha'$ and $\alpha'^2$ are fixed up to an overall factor by the above constraint. In this paper, we are going to use the above strategy to find $S_n^{(1)}$ at order $\alpha'^3$ in the type II superstring   and in the Heterotic string theories.

\section{Riemann curvature couplings at order $\alpha'^3$}

It is known that the three-point functions at two momentum level in the  superstring and Heterotic string theories are reproduced by their corresponding supergravites which have the following graviton and dilaton couplings:
\begin{align}
S_0=-\frac{2}{\kappa^2}\int d^{d+1}x e^{-2\Phi}\sqrt{-G}\,  \left(R + 4\nabla_{\alpha}\Phi \nabla^{\alpha}\Phi\right)\,.\labell{eq.1.110}
\end{align}
This action is invariant under the Buscher rules. There is no higher momentum corrections to the three-point functions in type II superstring theories, however, there are four momentum corrections to the three-point functions in the Heterotic string theory which are reproduced by the following   effective action  when there is no B-field:
\beqa
 S_1&=&\frac{-2}{\kappa}\alpha'\int d^{d+1}x e^{-2\Phi}\sqrt{-G}\Big( b_1 R_{\alpha \beta \gamma \delta} R^{\alpha \beta \gamma \delta}+ b_2 R_{\alpha \beta} R^{\alpha \beta} + b_3 R^2  \nonumber\\
&& +   b_4 R_{\alpha \beta} \nabla^{\alpha}\Phi \nabla^{\beta}\Phi+  b_5 R \nabla_{\alpha}\Phi \nabla^{\alpha}\Phi +b_6 R \nabla_{\alpha}\nabla^{\alpha}\Phi +   b_7 \nabla_{\alpha}\nabla^{\alpha}\Phi \nabla_{\beta}\nabla^{\beta}\Phi \nn \\ 
&&+ b_8 \nabla_{\alpha}\Phi \nabla^{\alpha}\Phi \nabla_{\beta}\nabla^{\beta}\Phi   -2 (8 b_{3} - 2 b_{5} - 4 b_{6} + 2 b_{7} + b_{8}) \nabla_{\alpha}\Phi \nabla^{\alpha}\Phi \nabla_{\beta}\Phi \nabla^{\beta}\Phi\Big)\labell{finalS10}
\eeqa
where $b_1=1/8$ for the Heterotic theory and $b_1=0$ for the type II superstring theories \cite{Metsaev:1986yb}. The couplings with coefficients $b_2,\cdots, b_8$ which belong to the $S_n^{(2)}$ part, are not fixed by the S-matrix elements.  They are changed under field redefinitions.   The form of effective action at the higher orders of $\alpha'$ depend on the form of   these couplings, so we   keep these terms in the effective action. It has been shown in \cite{Razaghian:2017okr} that the above action is invariant under T-duality at order $\alpha'$.

There is no six-momentum and higher corrections to the three-point functions in both type II and Heterotic    theories, hence, the higher derivative corrections to the above actions belonging to $S_n^{(1)}$ part of the effective action, must have at least four curvatures.   Using the cyclic symmetry for the Riemann curvature, one finds there are only seven  such independent couplings, \ie 
\beqa
S_3&=&-\frac{2}{\kappa^2}\int d^{d+1}x e^{-2\Phi}\sqrt{-G}\bigg[d_1 R_{\alpha \beta}{}^{\zeta \eta} R^{\alpha \beta \gamma \delta} R_{\gamma \zeta}{}^{\theta \iota} R_{\delta \eta \theta \iota} + d_2 R_{\alpha}{}^{\zeta}{}_{\gamma}{}^{\eta} R^{\alpha \beta \gamma \delta} R_{\beta}{}^{\theta}{}_{\zeta}{}^{\iota} R_{\delta \theta \eta \iota}\nn\\
&& + d_3 R_{\alpha \beta}{}^{\zeta \eta} R^{\alpha \beta \gamma \delta} R_{\gamma}{}^{\theta}{}_{\zeta}{}^{\iota} R_{\delta \theta \eta \iota} + d_4 R_{\alpha \beta}{}^{\zeta \eta} R^{\alpha \beta \gamma \delta} R_{\gamma \delta}{}^{\theta \iota} R_{\zeta \eta \theta \iota} + d_5 R_{\alpha \beta \gamma}{}^{\zeta} R^{\alpha \beta \gamma \delta} R_{\delta}{}^{\eta \theta \iota} R_{\zeta \eta \theta \iota}\nn\\
&& + d_6 R_{\alpha}{}^{\zeta}{}_{\gamma}{}^{\eta} R^{\alpha \beta \gamma \delta} R_{\beta}{}^{\theta}{}_{\delta}{}^{\iota} R_{\zeta \theta \eta \iota} + d_7 R_{\alpha \beta \gamma \delta} R^{\alpha \beta \gamma \delta} R_{\zeta \eta \theta \iota} R^{\zeta \eta \theta \iota}\bigg]\labell{RRRR}
\eeqa
where $d_1,d_2,\cdots, d_7$ are some unknown coefficients that we are going to find them by the T-duality constraint. 

Examining the structure of the terms with coefficients $d_5, d_7$, one realizes that they can be produced by the variation of  $ \sqrt{-G}R_{\mu\nu\alpha\beta}R^ {\mu\nu\alpha\beta}$ at order $\alpha'^2$, \ie
\beqa
 \delta(\alpha'\sqrt{-G}R_{\mu\nu\alpha\beta}R^ {\mu\nu\alpha\beta})&=&\alpha'^3\sqrt{-G}\Big(4R^{\gamma\alpha}R^{\delta}{}_{\alpha}+\frac{1}{2}G^{\gamma\delta}R_{\alpha\beta\mu\nu}R^{\alpha\beta\mu\nu}-4R^{\alpha\beta}R^{\gamma}{}_{\alpha}{}^{\delta}{}_{\beta}\nn\\&&-2R^{\gamma\alpha\beta\mu}R^{\delta}{}_{\alpha\beta\mu}-4\nabla_{\alpha}\nabla^{\alpha}R^{\gamma\delta}+2\nabla^{\delta}\nabla^{\gamma}R\Big)\delta G^{(2)}_{\gamma\delta}
\eeqa
 For appropriate   variation $\delta G^{(2)}_{\gamma\delta}$, the second  and the fourth terms  produce the terms with the coefficients $d_5,d_7$. So the coefficients $d_5,d_7$ in \reef{RRRR} are changed under field redefinitions in the Heterotic theory, whereas these terms do not change under the field redefinitions in type II superstring theory because this theory does not have the  Riemann squared coupling. Hence, these terms in the Heterotic theory belong to the $S_n^{(2)}$ part which can be set to zero for a specific field variables. Whereas in type II theory, they belong to $S_n^{(1)}$ part which should be fixed by the T-duality constraint. Since we are going to compare the couplings that the T-duality constraint produces with the couplings that the S-matrix method produces for which the field variables are not those that correspond to zero $d_5,d_7$ couplings, we keep these terms in both type II and Heterotic theories and let the T-duality fixes them.  

To study the T-duality transformation of the   couplings \reef{RRRR}, one should first reduce the  10-dimensional action to the 9-dimensional action. For the case that  B-field is zero, the T-duality transformations are   consistent for   diagonal metric. So we consider the reduction of metric as 
$G_{\mu\nu}= {\rm diag}(g_{ab},e^{2\sigma})$ where $g_{ab}$ is the $d$-dimensional metric.
 This reduction of metric produces the following reductions for the  different components of the Riemann curvature:
\begin{align}
&R_{abcd}= \tilde{R}_{abcd}\nn\\
&R_{abcy}=0\nn\\
& R_{ayby}= e^{2 \sigma} \left(-\tilde{\nabla}_{a}\sigma \tilde{\nabla}_{b}\sigma - \tilde{\nabla}_{b}\tilde{\nabla}_{a}\sigma\right)\labell{red}
\end{align}
 where we have  assumed that the fields are independent of the killing coordinate $ y $.  The tilde sign over the covariant derivatives and curvature means the metric in them is the $d$-dimensional metric $g_{ab}$.  Using the Mathematica package ``xAct'' \cite{Nutma:2013zea}, one can   separate the indices in \reef{RRRR} to the $d$-dimensional indices $a,b,c,\cdots$ and the killing $y$-index. Then using the reduction \reef{red}, one finds the following reduction for the action \reef{RRRR}: 
 \beqa
\bS_3&=&-\frac{2}{\kappa^2}\int d^{d}x e^{-2P}\sqrt{-g}\bigg[ d_1 \tilde{R}_{ab}{}^{ei} \tilde{R}^{abcd} \tilde{R}_{ce}{}^{jk} \tilde{R}_{dijk} + d_2 \tilde{R}_{a}{}^{e}{}_{c}{}^{i} \tilde{R}^{abcd} \tilde{R}_{b}{}^{j}{}_{e}{}^{k} \tilde{R}_{djik}\nn\\&& + d_3 \tilde{R}_{ab}{}^{ei} \tilde{R}^{abcd} \tilde{R}_{c}{}^{j}{}_{e}{}^{k} \tilde{R}_{djik} + d_4 \tilde{R}_{ab}{}^{ei} \tilde{R}^{abcd} \tilde{R}_{cd}{}^{jk} \tilde{R}_{eijk} 
  + d_5 \tilde{R}_{abc}{}^{e} \tilde{R}^{abcd} \tilde{R}_{d}{}^{ijk} \tilde{R}_{eijk} \nn\\&&+ d_6 \tilde{R}_{a}{}^{e}{}_{c}{}^{i} \tilde{R}^{abcd} \tilde{R}_{b}{}^{j}{}_{d}{}^{k} \tilde{R}_{ejik} + d_7 \tilde{R}_{abcd} \tilde{R}^{abcd} \tilde{R}_{eijk} \tilde{R}^{eijk}  
+ 8 d_7 \tilde{R}_{cdei} \tilde{R}^{cdei} \tilde{\nabla}_{a}\sigma \tilde{\nabla}^{a}\sigma \tilde{\nabla}_{b}\sigma \tilde{\nabla}^{b}\sigma \nn\\&&+ 16 d_7 \tilde{R}_{cdei} \tilde{R}^{cdei} \tilde{\nabla}^{a}\sigma \tilde{\nabla}_{b}\tilde{\nabla}_{a}\sigma \tilde{\nabla}^{b}\sigma + 8 d_7 \tilde{R}_{cdei} \tilde{R}^{cdei} \tilde{\nabla}_{b}\tilde{\nabla}_{a}\sigma \tilde{\nabla}^{b}\tilde{\nabla}^{a}\sigma  
\nn\\&&+ \tfrac{8}{3} d_5 \tilde{R}_{b}{}^{dei} \tilde{R}_{cdei} \tilde{\nabla}_{a}\sigma \tilde{\nabla}^{a}\sigma \tilde{\nabla}^{b}\sigma \tilde{\nabla}^{c}\sigma + \tfrac{8}{3} d_5 \tilde{R}_{b}{}^{dei} \tilde{R}_{cedi} \tilde{\nabla}_{a}\sigma \tilde{\nabla}^{a}\sigma \tilde{\nabla}^{b}\sigma \tilde{\nabla}^{c}\sigma \nn\\&&+ \tfrac{16}{3} d_5 \tilde{R}_{b}{}^{dei} \tilde{R}_{cdei} \tilde{\nabla}^{a}\sigma \tilde{\nabla}^{b}\sigma \tilde{\nabla}^{c}\tilde{\nabla}_{a}\sigma  
+ \tfrac{16}{3} d_5 \tilde{R}_{b}{}^{dei} \tilde{R}_{cedi} \tilde{\nabla}^{a}\sigma \tilde{\nabla}^{b}\sigma \tilde{\nabla}^{c}\tilde{\nabla}_{a}\sigma \nn\\&&+ \tfrac{8}{3} d_5 \tilde{R}_{b}{}^{dei} \tilde{R}_{cdei} \tilde{\nabla}^{b}\tilde{\nabla}^{a}\sigma \tilde{\nabla}^{c}\tilde{\nabla}_{a}\sigma + \tfrac{8}{3} d_5 \tilde{R}_{b}{}^{dei} \tilde{R}_{cedi} \tilde{\nabla}^{b}\tilde{\nabla}^{a}\sigma \tilde{\nabla}^{c}\tilde{\nabla}_{a}\sigma 
 \nn\\&&+ 2 (d_2 + 2 d_6) \tilde{R}_{a}{}^{e}{}_{b}{}^{i} \tilde{R}_{cedi} \tilde{\nabla}^{a}\sigma \tilde{\nabla}^{b}\sigma \tilde{\nabla}^{c}\sigma \tilde{\nabla}^{d}\sigma - 4 (d_2 + d_3) \tilde{R}_{bdce} \tilde{\nabla}^{b}\tilde{\nabla}^{a}\sigma \tilde{\nabla}^{c}\tilde{\nabla}_{a}\sigma \tilde{\nabla}^{e}\tilde{\nabla}^{d}\sigma 
 \nn\\&&+ 2 (4 d_1 + d_2 + 8 d_4 + 4 d_5 + 2 d_6 + 8 d_7) \tilde{\nabla}_{a}\sigma \tilde{\nabla}^{a}\sigma \tilde{\nabla}_{b}\sigma \tilde{\nabla}^{b}\sigma \tilde{\nabla}_{c}\sigma \tilde{\nabla}^{c}\sigma \tilde{\nabla}_{d}\sigma \tilde{\nabla}^{d}\sigma 
 \nn\\&&+ 8 (4 d_1 + d_2 + 8 d_4 + 4 d_5 + 2 d_6 + 8 d_7) \tilde{\nabla}_{a}\sigma \tilde{\nabla}^{a}\sigma \tilde{\nabla}_{b}\sigma \tilde{\nabla}^{b}\sigma \tilde{\nabla}^{c}\sigma \tilde{\nabla}_{d}\tilde{\nabla}_{c}\sigma \tilde{\nabla}^{d}\sigma  
\nn\\&&+ 2 (8 d_1 + 3 d_2 + 16 d_4 + 12 d_5 + 6 d_6 + 32 d_7) \tilde{\nabla}^{a}\sigma \tilde{\nabla}_{b}\tilde{\nabla}_{a}\sigma \tilde{\nabla}^{b}\sigma \tilde{\nabla}^{c}\sigma \tilde{\nabla}_{d}\tilde{\nabla}_{c}\sigma \tilde{\nabla}^{d}\sigma  
\nn\\&&+ 4 \bigl(8 d_1 + d_2 + 2 (8 d_4 + 2 d_5 + d_6)\bigr) \tilde{\nabla}_{a}\sigma \tilde{\nabla}^{a}\sigma \tilde{\nabla}^{b}\sigma \tilde{\nabla}^{c}\sigma \tilde{\nabla}_{d}\tilde{\nabla}_{c}\sigma \tilde{\nabla}^{d}\tilde{\nabla}_{b}\sigma 
\nn\\&&+ 4 \bigl(8 d_1 + d_2 + 2 (8 d_4 + 2 d_5 + d_6)\bigr) \tilde{\nabla}^{a}\sigma \tilde{\nabla}^{b}\sigma \tilde{\nabla}^{c}\tilde{\nabla}_{a}\sigma \tilde{\nabla}_{d}\tilde{\nabla}_{c}\sigma \tilde{\nabla}^{d}\tilde{\nabla}_{b}\sigma  
\nn\\&&+ \bigl(8 d_1 + d_2 + 2 (8 d_4 + 2 d_5 + d_6)\bigr) \tilde{\nabla}^{b}\tilde{\nabla}^{a}\sigma \tilde{\nabla}^{c}\tilde{\nabla}_{a}\sigma \tilde{\nabla}_{d}\tilde{\nabla}_{c}\sigma \tilde{\nabla}^{d}\tilde{\nabla}_{b}\sigma  
\nn\\&&+ \tfrac{4}{9} (d_2 + 2 d_3) \tilde{R}_{ac}{}^{ei} \tilde{R}_{bdei} \tilde{\nabla}^{a}\sigma \tilde{\nabla}^{b}\sigma \tilde{\nabla}^{d}\tilde{\nabla}^{c}\sigma + \tfrac{8}{9} (d_2 + 2 d_3) \tilde{R}_{ac}{}^{ei} \tilde{R}_{bedi} \tilde{\nabla}^{a}\sigma \tilde{\nabla}^{b}\sigma \tilde{\nabla}^{d}\tilde{\nabla}^{c}\sigma
\nn\\&&+ \tfrac{4}{9} (5 d_2 + d_3) \tilde{R}_{a}{}^{e}{}_{c}{}^{i} \tilde{R}_{bedi} \tilde{\nabla}^{a}\sigma \tilde{\nabla}^{b}\sigma \tilde{\nabla}^{d}\tilde{\nabla}^{c}\sigma + \tfrac{4}{9} (4 d_2 -  d_3) \tilde{R}_{a}{}^{e}{}_{c}{}^{i} \tilde{R}_{bide} \tilde{\nabla}^{a}\sigma \tilde{\nabla}^{b}\sigma \tilde{\nabla}^{d}\tilde{\nabla}^{c}\sigma  
\nn\\&&+ 8 d_6 \tilde{R}_{a}{}^{e}{}_{b}{}^{i} \tilde{R}_{cedi} \tilde{\nabla}^{a}\sigma \tilde{\nabla}^{b}\sigma \tilde{\nabla}^{d}\tilde{\nabla}^{c}\sigma + \tfrac{2}{9} (d_2 + 2 d_3) \tilde{R}_{ac}{}^{ei} \tilde{R}_{bdei} \tilde{\nabla}^{b}\tilde{\nabla}^{a}\sigma \tilde{\nabla}^{d}\tilde{\nabla}^{c}\sigma  
\nn\\&&+ \tfrac{4}{9} (d_2 + 2 d_3) \tilde{R}_{ac}{}^{ei} \tilde{R}_{bedi} \tilde{\nabla}^{b}\tilde{\nabla}^{a}\sigma \tilde{\nabla}^{d}\tilde{\nabla}^{c}\sigma + \tfrac{2}{9} (5 d_2 + d_3) \tilde{R}_{a}{}^{e}{}_{c}{}^{i} \tilde{R}_{bedi} \tilde{\nabla}^{b}\tilde{\nabla}^{a}\sigma \tilde{\nabla}^{d}\tilde{\nabla}^{c}\sigma  
\nn\\&&+ \tfrac{2}{9} (4 d_2 -  d_3) \tilde{R}_{a}{}^{e}{}_{c}{}^{i} \tilde{R}_{bide} \tilde{\nabla}^{b}\tilde{\nabla}^{a}\sigma \tilde{\nabla}^{d}\tilde{\nabla}^{c}\sigma + 4 d_6 \tilde{R}_{a}{}^{e}{}_{b}{}^{i} \tilde{R}_{cedi} \tilde{\nabla}^{b}\tilde{\nabla}^{a}\sigma \tilde{\nabla}^{d}\tilde{\nabla}^{c}\sigma  
\nn\\&&+ 2 \bigl(d_2 + 2 (2 d_5 + d_6 + 8 d_7)\bigr) \tilde{\nabla}_{a}\sigma \tilde{\nabla}^{a}\sigma \tilde{\nabla}_{b}\sigma \tilde{\nabla}^{b}\sigma \tilde{\nabla}_{d}\tilde{\nabla}_{c}\sigma \tilde{\nabla}^{d}\tilde{\nabla}^{c}\sigma  
\nn\\&&+ 4 \bigl(d_2 + 2 (2 d_5 + d_6 + 8 d_7)\bigr) \tilde{\nabla}^{a}\sigma \tilde{\nabla}_{b}\tilde{\nabla}_{a}\sigma \tilde{\nabla}^{b}\sigma \tilde{\nabla}_{d}\tilde{\nabla}_{c}\sigma \tilde{\nabla}^{d}\tilde{\nabla}^{c}\sigma 
\nn\\&&+ \bigl(d_2 + 2 (2 d_5 + d_6 + 8 d_7)\bigr) \tilde{\nabla}_{b}\tilde{\nabla}_{a}\sigma \tilde{\nabla}^{b}\tilde{\nabla}^{a}\sigma \tilde{\nabla}_{d}\tilde{\nabla}_{c}\sigma \tilde{\nabla}^{d}\tilde{\nabla}^{c}\sigma  
\nn\\&&- 4 (d_2 + d_3) \tilde{R}_{adbe} \tilde{\nabla}^{a}\sigma \tilde{\nabla}^{b}\sigma \tilde{\nabla}^{d}\tilde{\nabla}^{c}\sigma \tilde{\nabla}^{e}\tilde{\nabla}_{c}\sigma - 4 (d_2 + d_3) \tilde{R}_{bdce} \tilde{\nabla}_{a}\sigma \tilde{\nabla}^{a}\sigma \tilde{\nabla}^{b}\sigma \tilde{\nabla}^{c}\sigma \tilde{\nabla}^{e}\tilde{\nabla}^{d}\sigma 
 \nn\\&&- 8 (d_2 + d_3) \tilde{R}_{bdce} \tilde{\nabla}^{a}\sigma \tilde{\nabla}^{b}\sigma \tilde{\nabla}^{c}\tilde{\nabla}_{a}\sigma \tilde{\nabla}^{e}\tilde{\nabla}^{d}\sigma\bigg]
 \eeqa
where the $d$-dimensional dilaton is $P=\Phi-\sigma/2$. The transformation of $\bS_3$ under the Buscher rules is constrained to be zero, \ie $\delta \bS_3=0$, up to the $d$-dimensional field redefinitions. Under the Buscher rules, 
the terms in $\bS_3$ with odd number of $\sigma$ are survived, \ie 
\beqa
\delta \bS_3&\!\!\!\!=\!\!\!\!&-\frac{2}{\kappa^2}\int d^{d}x e^{-2P}\sqrt{-g}\bigg[32 d_7 \tilde{R}_{cdei} \tilde{R}^{cdei} \tilde{\nabla}^{a}\sigma \tilde{\nabla}_{b}\tilde{\nabla}_{a}\sigma \tilde{\nabla}^{b}\sigma + \tfrac{32}{3} d_5 \tilde{R}_{b}{}^{dei} \tilde{R}_{cdei} \tilde{\nabla}^{a}\sigma \tilde{\nabla}^{b}\sigma \tilde{\nabla}^{c}\tilde{\nabla}_{a}\sigma \nn\\&& + \tfrac{32}{3} d_5 \tilde{R}_{b}{}^{dei} \tilde{R}_{cedi} \tilde{\nabla}^{a}\sigma \tilde{\nabla}^{b}\sigma \tilde{\nabla}^{c}\tilde{\nabla}_{a}\sigma - 8 (d_2 + d_3) \tilde{R}_{bdce} \tilde{\nabla}^{b}\tilde{\nabla}^{a}\sigma \tilde{\nabla}^{c}\tilde{\nabla}_{a}\sigma \tilde{\nabla}^{e}\tilde{\nabla}^{d}\sigma \nn\\&&+ 16 (4 (d_1+ 2 d_4) + d_2  + 4 d_5 + 2 d_6 + 8 d_7) \tilde{\nabla}_{a}\sigma \tilde{\nabla}^{a}\sigma \tilde{\nabla}_{b}\sigma \tilde{\nabla}^{b}\sigma \tilde{\nabla}^{c}\sigma \tilde{\nabla}_{d}\tilde{\nabla}_{c}\sigma \tilde{\nabla}^{d}\sigma \nn\\&& + 8 \bigl(8( d_1 +2d_4)+ d_2 + 2 ( 2 d_5 + d_6)\bigr) \tilde{\nabla}^{a}\sigma \tilde{\nabla}^{b}\sigma \tilde{\nabla}^{c}\tilde{\nabla}_{a}\sigma \tilde{\nabla}_{d}\tilde{\nabla}_{c}\sigma \tilde{\nabla}^{d}\tilde{\nabla}_{b}\sigma \nn\\&& + \tfrac{8}{9} (d_2 + 2 d_3) \tilde{R}_{ac}{}^{ei} \tilde{R}_{bdei} \tilde{\nabla}^{a}\sigma \tilde{\nabla}^{b}\sigma \tilde{\nabla}^{d}\tilde{\nabla}^{c}\sigma + \tfrac{16}{9} (d_2 + 2 d_3) \tilde{R}_{ac}{}^{ei} \tilde{R}_{bedi} \tilde{\nabla}^{a}\sigma \tilde{\nabla}^{b}\sigma \tilde{\nabla}^{d}\tilde{\nabla}^{c}\sigma \nn\\&& + \tfrac{8}{9} (5 d_2 + d_3) \tilde{R}_{a}{}^{e}{}_{c}{}^{i} \tilde{R}_{bedi} \tilde{\nabla}^{a}\sigma \tilde{\nabla}^{b}\sigma \tilde{\nabla}^{d}\tilde{\nabla}^{c}\sigma + \tfrac{8}{9} (4 d_2 -  d_3) \tilde{R}_{a}{}^{e}{}_{c}{}^{i} \tilde{R}_{bide} \tilde{\nabla}^{a}\sigma \tilde{\nabla}^{b}\sigma \tilde{\nabla}^{d}\tilde{\nabla}^{c}\sigma \nn\\&&+ 16 d_6 \tilde{R}_{a}{}^{e}{}_{b}{}^{i} \tilde{R}_{cedi} \tilde{\nabla}^{a}\sigma \tilde{\nabla}^{b}\sigma \tilde{\nabla}^{d}\tilde{\nabla}^{c}\sigma - 8 (d_2 + d_3) \tilde{R}_{bdce} \tilde{\nabla}_{a}\sigma \tilde{\nabla}^{a}\sigma \tilde{\nabla}^{b}\sigma \tilde{\nabla}^{c}\sigma \tilde{\nabla}^{e}\tilde{\nabla}^{d}\sigma \nn\\&& + 8 \bigl(d_2 + 2 (2 d_5 + d_6 + 8 d_7)\bigr) \tilde{\nabla}^{a}\sigma \tilde{\nabla}_{b}\tilde{\nabla}_{a}\sigma \tilde{\nabla}^{b}\sigma \tilde{\nabla}_{d}\tilde{\nabla}_{c}\sigma \tilde{\nabla}^{d}\tilde{\nabla}^{c}\sigma\bigg]\,=\,0\labell{sons}
\eeqa
As can be seen, the coefficients $d_1$, $d_4$ appear only through the combination $d_1+2d_4$. This is resulted from the reduction \reef{red} which has the simple form for the   case that   metric is diagonal.  So the T-duality for the case that   metric is diagonal can not fix the coefficients $d_1$, $d_4$ separately. However, all other coefficients appear in different forms in different terms. So we expect the T-duality fix them separately.
 
Since the   constrain \reef{sons} is on the action, one is free to add to the Lagrangian all total covariant derivative terms at order $\alpha'^3$ which have odd number of $\sigma$. Using the ``xAct'', it is very simple to construct all such total derivative terms. One should first write all contraction of curvature, covariant derivatives of $\sigma$ and covariant derivatives of $P$ which have odd number
of $\sigma$, at seven derivative order with one free index. We choose the coefficient of each term to be   arbitrary. Then we multiply them with the $d$-dimensional dilaton factor $e^{-2P}$. We call the resulting vector to be $J^a$. Then taking  a covariant derivative on $J^a$, \ie $\nabla_a J^a$, one finds all $d$-dimensional total derivative terms.  If one adds to the constraint  \reef{sons} all the $d$-dimensional total derivative terms, one would find the wrong result that $d_1+2d_4=d_2=d_3=d_5=d_6=d_7=0$. Therefore, we have to take into account  the $d$-dimensional field redefinitions as well.

To construct the $d$-dimensional field redefinitions, one should first reduce the lower $\alpha'$-order $D$-dimensional actions  \reef{eq.1.110} and \reef{finalS10} to the $d$ dimensions. 
Then one should consider the transformation of the resulting actions under the following field redefinitions:
\beqa
\sigma&\rightarrow &-\sigma+\delta\sigma\nn\\
P&\rightarrow &P+\delta P\nn\\
g_{ab}&\rightarrow &g_{ab}+\delta g_{ab}\labell{per}
\eeqa
The corrections to the Buscher rules, \ie $\delta\sigma$, $\delta P$ and $\delta g_{ab}$, for the type II theory begin at order $\alpha'^3$ because there is no effective actions at orders $\alpha'$ and $\alpha'^2$. In the Heterotic theory the corrections begin at order $\alpha'$. So, let use consider each case separately.

\subsection{Couplings in  Type II supergravity }

It is known that the Type IIA theory transforms to the Type IIB theory under the T-duality transformation \cite{Dai:1989ua,Dine:1989vu}. The effective actions of these theories, however, are identical in the NS-NS sector. As a result, the NS-NS couplings at any order of $\alpha'$ must be invariant under the T-duality transformation. The effective actions of these theories  at the leading order of $\alpha'$ are invariant under the Buscher rules \cite{Bergshoeff:1995as}, however, the $\alpha'^3$-corrections to these couplings  are not invariant under the Buscher  rules unless one extends them by some  $\alpha'^3$-corrections \ie 
\beqa
\sigma&\rightarrow &-\sigma+\alpha'^3\delta\sigma^{(3)}\nn\\
P&\rightarrow &P+\alpha'^3\delta P^{(3)}\nn\\
g_{ab}&\rightarrow &g_{ab}+\alpha'^3\delta g^{(3)}_{ab}\labell{per3}
\eeqa
One should replace \reef{per3} in the reduction of \reef{eq.1.110} which is  
\beqa
\bS_0&=&-\frac{2}{\kappa^2}\int d^{d}x e^{-2P} \sqrt{-g}\, \Big(\tilde{R}  + 4 \tilde{\nabla}_{a}P \tilde{\nabla}^{a}P  - \tilde{\nabla}_{a}\sigma \tilde{\nabla}^{a}\sigma\Big) 
\eeqa
and keep  terms linear in the variations. Up to some total derivative terms, the variations  $\delta \sigma^{(3)} ,\ \delta P^{(3)} ,  \delta g^{(3)} _{ab}$ produce the following variation for $\bS_0$:
 \beqa
{  \delta \bS_0}&=&\bS_0\Big(-\sigma+\alpha'^3\delta\sigma^{(3)},P+\alpha'^3\delta P^{(3)},g_{ab}+\alpha'^3\delta g^{(3)}_{ab}\Big)-\bS_0(\sigma,P,g_{ab})\nn\\
&=&\frac{2\alpha'^3}{\kappa^2}\int d^{d}x e^{-2P} \sqrt{-g}\bigg[2( \tilde{\nabla}_{a}\tilde{\nabla}^{a}\sigma - 2 \tilde{\nabla}_{a}\sigma \tilde{\nabla}^{a}P)\delta\sigma^{(3)}\nn\\
&&+\Big( \tilde{R}^{ab}  +2 \tilde{\nabla}^{a} \tilde{\nabla}^{b}P -\tilde{\nabla}^{a}\sigma \tilde{\nabla}^{b}\sigma -\tfrac{1}{2} g^{ab}  (\tilde{R}+  4\tilde{\nabla}_{c} \tilde{\nabla}^{c}P   - 4    \tilde{\nabla}_{c}P \tilde{\nabla}^{c}P-  
 \tilde{\nabla}_{c}\sigma \tilde{\nabla}^{c}\sigma)\Big)\delta g^{(3)}_{ab} \nonumber \\ 
&& + 2 (\tilde{R}+4 \tilde{\nabla}_{a}\tilde{\nabla}^{a}P -4 \tilde{\nabla}_{a}P \tilde{\nabla}^{a}P - \tilde{\nabla}_{a}\sigma \tilde{\nabla}^{a}\sigma)\delta P^{(3)} \bigg]+\cdots\labell{eq.2.2.10}
\eeqa
where dots represent terms at higher orders of $\alpha'$ in which we are not interested. In order to produce   couplings at order $\alpha'^3$, the variations  $\delta \sigma^{(3)} ,\ \delta P^{(3)} ,  \delta g^{(3)} _{ab}$ should be all contractions of the $d$-dimensional fields  at six derivative level with unknown coefficients. To produce the field redefinitions  with odd number of $\sigma$ as in \reef{sons}, one should consider terms in $\delta \sigma^{(3)}$ that have even number of $\sigma$, and terms in $\delta P^{(3)},  \delta g^{(3)}_{ab}$ that have odd number of $\sigma$.  
Adding these field redefinition terms   as well as all total derivative terms  to the constraint \reef{sons}, and writing them in terms of independent couplings, one finds many algebraic equations involving the $d$-coefficients, the coefficients of the total derivative terms, and the coefficients of the variations $\delta \sigma^{(3)} ,\ \delta P^{(3)} ,  \delta g^{(3)} _{ab}$. Solving these equations, one finds   many coefficients in the  corrections  $\delta \sigma^{(3)} ,\ \delta P^{(3)} ,  \delta g^{(3)} _{ab}$ are not fixed, and the remaining  coefficients are fixed in terms of the unfixed coefficients and the $d$-coefficients.  The equations  for zero  $\delta \sigma^{(3)} ,\ \delta P^{(3)} ,  \delta g^{(3)} _{ab}$ fix the effective action to be zero, hence, the non-zero effective action forces the Buscher rules to receive $\alpha'^3$-corrections. The equations for non-zero  $\delta \sigma^{(3)} ,\ \delta P^{(3)} ,  \delta g^{(3)} _{ab}$, however,   fix the $d$-coefficients in the effective action \reef{RRRR} up to an overall factor, \ie
\beqa
 d_1+2d_4=-\frac{d_2}{4}&;&d_3=d_5=d_6=d_7=0
\eeqa
The only unknown coefficients is $d_2$.  Note   that, since the coefficients $d_5,d_7$ are not changed under the field redefinitions in type II theory, they are fixed by the T-duality constraint. 

As we have pointed out before, the coefficients $d_1$ and $d_4$   appear as one coefficient $d_1+2d_4$. We expect the coefficient $d_4$ to be fixed by the T-duality if one extends the present calculations in which there is no  $B$-field, to the calculations in the presence of  B-field  which  we leave it for the future works. If we choose it to be zero, the effective action \reef{RRRR} then becomes
\beqa
S_3&=&-\frac{2d_1}{\kappa^2}\int d^{10}x e^{-2\Phi}\sqrt{-G}\bigg[ R_{\alpha \beta}{}^{\zeta \eta} R^{\alpha \beta \gamma \delta} R_{\gamma \zeta}{}^{\theta \iota} R_{\delta \eta \theta \iota} -4 R_{\alpha}{}^{\zeta}{}_{\gamma}{}^{\eta} R^{\alpha \beta \gamma \delta} R_{\beta}{}^{\theta}{}_{\zeta}{}^{\iota} R_{\delta \theta \eta \iota} \bigg]\labell{II}
\eeqa
In type II theory, using the KLT relation  between the scattering amplitudes of the closed strings  and the scattering amplitudes of   open strings    \cite{Kawai:1985xq}, one expects the closed string couplings to be written as  produce of two open string couplings. Using   the   $t_8$ tensor which was first  defined in \cite{Schwarz:1982jn}, \ie the contraction of $t_8$ with   four arbitrary antisymmetric tensors $M^1,\,\cdots, M^4$  is  
\beqa
t^{\alpha\beta\gamma\delta\mu\nu\rho\sigma}M^1_{\alpha\beta}M^2_{\gamma\delta}M^3_{\mu\nu}M^4_{\rho\sigma}&\!\!\!\!\!=\!\!\!\!\!&8(\tr M^1M^2M^3M^4+\tr M^1M^3M^2M^4+\tr M^1M^3M^4M^2)\labell{t8}\\
&&-2(\tr M^1M^2\tr M^3M^4+\tr M^1M^3\tr M^2M^4+\tr M^1M^4\tr M^2M^3)\nn
\eeqa
and the Levi-Civita tensor $\epsilon_{10}$, the couplings \reef{II} can be written as   the following expression:
\beqa
S_3&=&- \frac{2d_1}{3.2^7\kappa^2} \int d^{10}x e^{-2\Phi} \sqrt{-G}(t_8t_8R^4+\frac{1}{8}\eps_{10}\eps_{10}R^4 )\labell{Y1}
\eeqa
  For $d_1=\alpha'^3\z(3)/2^7$, this is exactly the $R^4$-correction to the supergravity that was  first found   from the sphere-level four-graviton scattering amplitude   \cite{ Schwarz:1982jn,Gross:1986iv} as well as from the $\sigma$-model beta function approach \cite{Grisaru:1986vi,Freeman:1986zh}.    
The Riemann curvature couplings given by $t_8t_8R^4$, \ie
\beqa
t_8t_8R^4&\equiv&t^{\mu_1\cdots \mu_8}t^{\nu_1\cdots \nu_8}R_{\mu_1\mu_2}{}_{\nu_1\nu_2}R_{\mu_3\mu_4}{}_{\nu_3\nu_4}R_{\mu_5\mu_6}{}_{\nu_5\nu_6}R_{\mu_7\mu_8}{}_{\nu_7\nu_8}\nonumber\\
&=&3\cdot 2^7\bigg[R_{\alpha \beta  \gamma  \delta  }R_{\beta  \mu  \delta  \rho} R_{\mu  \nu   \sigma \gamma  }R_{\nu  \alpha  \rho \sigma} +\frac{1}{2}R_{\alpha \beta  \gamma  \delta  }R_{\beta  \mu  \delta  \rho }R_{\mu  \nu  \rho \sigma }R_{\nu  \alpha \sigma \gamma  }\labell{Y3}\\
&&  -\frac{1}{2}R_{\alpha \beta  \gamma  \delta  }R_{\beta  \mu  \gamma  \delta  }R_{\mu  \nu  \rho \sigma }R_{\nu  \alpha \rho \sigma } -\frac{1}{4}R_{\alpha \beta  \gamma  \delta  }R_{\beta  \mu  \rho \sigma }R_{\mu  \nu  }{}_{\gamma  \delta  }R_{\nu  \alpha }{}_{\rho \sigma }\nonumber  \\
&& +\frac{1}{16}R_{\alpha \beta  \gamma  \delta  }R_{\beta  \alpha \rho \sigma }R_{\mu  \nu  \gamma  \delta  }R_{\nu  \mu  \rho \sigma }+\frac{1}{32}R_{\alpha \beta  \gamma  \delta  }R_{\beta  \alpha \gamma  \delta  }R_{\mu  \nu  \rho \sigma }R_{\nu  \mu  \rho \sigma }\bigg]\nonumber
\eeqa
have nonzero contribution at four-graviton level, so they were found from the sphere-level S-matrix element of four graviton vertex operators \cite{ Schwarz:1982jn,Gross:1986iv}, whereas the couplings given by $\eps_{10}\eps_{10}R^4$ whose Riemann curvature couplings are
\beqa
\frac{1}{8}\eps_{10}\eps_{10}R^4&=&3\cdot 2^7\bigg[-R_{\alpha  \beta   \gamma   \delta  } R_{\rho  \sigma  \beta   \mu  } R_{\delta   \mu   \sigma  \nu  } R_{\gamma   \nu   \alpha  \rho }+R_{\alpha  \beta   \gamma   \delta  } R_{\rho  \sigma  \alpha  \beta  } R_{\delta   \mu   \sigma  \nu  } R_{\gamma   \nu   \rho  \mu  }\nonumber  \\
&&+\frac{1}{2} R_{\alpha  \beta   \gamma   \delta  } R_{\rho  \sigma  \mu   \nu  } R_{\gamma   \mu   \rho  \sigma } R_{\delta   \nu   \alpha  \beta  }-\frac{1}{2} R_{\alpha  \beta   \gamma   \delta  } R_{\rho  \sigma  \mu   \nu  } R_{\gamma   \mu   \alpha  \rho } R_{\delta   \nu   \beta   \sigma }\nonumber \\
&&-\frac{1}{16} R_{\alpha  \beta   \gamma   \delta  } R_{\gamma   \delta   \rho  \sigma } R_{\rho  \sigma  \mu   \nu  } R_{\mu   \nu   \alpha  \beta  }-\frac{1}{32} R_{\alpha  \beta   \gamma   \delta  } R_{\gamma   \delta   \alpha  \beta  } R_{\rho  \sigma  \mu   \nu  } R_{\mu   \nu   \rho  \sigma }\bigg]
\eeqa
have nonzero contribution at five-graviton level \cite{Zumino:1985dp}. However, the presence of this term in the tree-level effective action was first dictated by the $\sigma$-model beta function approach \cite{Grisaru:1986vi,Freeman:1986zh}. It has been shown in \cite{Garousi:2013tca} that the sphere-level scattering amplitude of five gravitons confirms the presence of $\eps_{10}\eps_{10}R^4$ in the tree-level effective action. It is interesting that the T-duality constrain could fix the presence of both terms in the effective action.

In the type IIB theory, one expects the couplings to be invariant under   S-duality as well. The invariance under the $SL(2,R)$ imposes the condition that the couplings in the Einstein frame must have no term with odd number of dilaton \cite{Garousi:2013qka}. If one transforms the T-duality invariant couplings \reef{II} to the Einstein frame, one would find couplings which have odd number of dilaton. On the other hand, because of the overall dilaton factor in the Einstein frame, one observers that  each total derivative term includes terms with odd and even number of dilatons. So the odd number of the dilatons in   transforming the  couplings \reef{II} to the Einstein frame, may be canceled by appropriate total derivative terms. We have checked that the terms which have odd number of dilaton can be canceled by adding some total derivative terms in the Einstein frame, \ie    the action \reef{II} is consistent with the S-duality.

\subsection{Couplings in  Heterotic supergravity }

In the Heterotic theory, the corrections to the Buscher rules begin at order $\alpha'$, \ie
\beqa
\sigma&\rightarrow &-\sigma+\alpha'\delta\sigma^{(1)}\nn\\
P&\rightarrow &P+\alpha'\delta P^{(1)}\nn\\
g_{ab}&\rightarrow &g_{ab}+\alpha'\delta g^{(1)}_{ab}\labell{per1}
\eeqa
where the corrections  are  parametrized by nine parameters
\beqa
\delta\sigma^{(1)} &=&A_1 \tilde{R} + A_2 \tilde{\nabla}_{a}\tilde{\nabla}^{a}P  + A_3 \tilde{\nabla}_{a}P \tilde{\nabla}^{a}P + A_4 \tilde{\nabla}_{a}\sigma \tilde{\nabla}^{a}\sigma \nonumber\\
\delta P^{(1)} &=&A_5 \tilde{\nabla}_{a}\tilde{\nabla}^{a}\sigma + A_6 \tilde{\nabla}_{a}\sigma \tilde{\nabla}^{a}P  \nonumber\\
\delta g_{ab}^{(1)}&=& A_7 (\tfrac{1}{2} \tilde{\nabla}_{a}\sigma \tilde{\nabla}_{b}P+ \tfrac{1}{2} \tilde{\nabla}_{a}P \tilde{\nabla}_{b}\sigma)   +g_{ab}\Big( A_8\tilde{\nabla}_{c}\tilde{\nabla}^{c}\sigma  
 + A_{9}  \tilde{\nabla}_{c}\sigma \tilde{\nabla}^{c}P \Big) \labell{eq.2.2.20}
\eeqa
We have   excluded the parameter corresponding to the  $d$-dimensional coordinate transformations, These corrections are required to make the $d$-dimensional  reduction of the  couplings \reef{finalS10}, \ie
\begin{align}
\bS_1=&-\frac{2}{\kappa^2}\alpha'\int  d^dx e^{-2P}\sqrt{-g}\Big(
 b_1 \tilde{R}_{abcd} \tilde{R}^{abcd}+b_2 \tilde{R}_{ab} \tilde{R}^{ab} + b_3 \tilde{R}^2  + b_6 \tilde{R} \tilde{\nabla}_{a}\tilde{\nabla}^{a}P  \nonumber \\ 
& + b_5 \tilde{R} \tilde{\nabla}_{a}P \tilde{\nabla}^{a}P + (b_5 + b_6) \tilde{R} \tilde{\nabla}_{a}\sigma \tilde{\nabla}^{a}P + \tfrac{1}{4} (-16 b_3 + b_5 + 2 b_6) \tilde{R} \tilde{\nabla}_{a}\sigma \tilde{\nabla}^{a}\sigma \nonumber \\ 
& - 2 b_2 \tilde{R}^{ab} \tilde{\nabla}_{b}\tilde{\nabla}_{a}\sigma + b_7 \tilde{\nabla}_{a}\tilde{\nabla}^{a}P \tilde{\nabla}_{b}\tilde{\nabla}^{b}P + b_8 \tilde{\nabla}_{a}P \tilde{\nabla}^{a}P \tilde{\nabla}_{b}\tilde{\nabla}^{b}P \nonumber \\ 
&+ (2 b_7  + b_8) \tilde{\nabla}_{a}\sigma \tilde{\nabla}^{a}P \tilde{\nabla}_{b}\tilde{\nabla}^{b}P + (-2 b_6 + b_7 + \tfrac{1}{4} b_8) \tilde{\nabla}_{a}\sigma \tilde{\nabla}^{a}\sigma \tilde{\nabla}_{b}\tilde{\nabla}^{b}P \nonumber \\ 
&  + (-2 b_6+ b_7) \tilde{\nabla}_{a}\tilde{\nabla}^{a}P \tilde{\nabla}_{b}\tilde{\nabla}^{b}\sigma + (b_2 + 4 b_3 -  b_6 + \tfrac{1}{4} b_7) \tilde{\nabla}_{a}\tilde{\nabla}^{a}\sigma \tilde{\nabla}_{b}\tilde{\nabla}^{b}\sigma \nonumber \\ 
& + \tfrac{1}{2} (-4 b_5 + b_8) \tilde{\nabla}_{a}P \tilde{\nabla}^{a}P \tilde{\nabla}_{b}\tilde{\nabla}^{b}\sigma + (-2 b_5 - 2 b_6 + b_7 + \tfrac{1}{2} b_8) \tilde{\nabla}_{a}\sigma \tilde{\nabla}^{a}P \tilde{\nabla}_{b}\tilde{\nabla}^{b}\sigma \nonumber \\ 
& + \tfrac{1}{8} (16 b_2 + 64 b_3 - 4 b_5 - 16 b_6 + 4 b_7 + b_8) \tilde{\nabla}_{a}\sigma \tilde{\nabla}^{a}\sigma \tilde{\nabla}_{b}\tilde{\nabla}^{b}\sigma + b_4 \tilde{R}_{ab} \tilde{\nabla}^{a}P \tilde{\nabla}^{b}P \nonumber \\ 
& - (2b_8+4b_7-8b_6-4b_5+16b_3) \tilde{\nabla}_{a}P \tilde{\nabla}^{a}P \tilde{\nabla}_{b}P \tilde{\nabla}^{b}P + b_4 \tilde{R}_{ab} \tilde{\nabla}^{a}P \tilde{\nabla}^{b}\sigma\nonumber \\ 
&+ (- b_4 -3b_3-b_8+8b_6+4b_5-16b_3) \tilde{\nabla}_{a}\sigma \tilde{\nabla}^{a}P \tilde{\nabla}_{b}\sigma \tilde{\nabla}^{b}P -  b_4 \tilde{\nabla}^{a}P \tilde{\nabla}_{b}\tilde{\nabla}_{a}\sigma \tilde{\nabla}^{b}P  \nonumber \\ 
& + \tfrac{1}{4} (-8 b_2 + b_4) \tilde{R}_{ab} \tilde{\nabla}^{a}\sigma \tilde{\nabla}^{b}\sigma + \tfrac{1}{2} (-  b_8 -4b_7+8b_6-16b_3 ) \tilde{\nabla}_{a}P \tilde{\nabla}^{a}P \tilde{\nabla}_{b}\sigma \tilde{\nabla}^{b}\sigma \nonumber \\ 
& + (- b_4 + 2 b_6 - b_7 - \tfrac{1}{4} b_8  ) \tilde{\nabla}_{a}\sigma \tilde{\nabla}^{a}P \tilde{\nabla}_{b}\sigma \tilde{\nabla}^{b}\sigma  + (8 b_1 + 2 b_2 -  \tfrac{1}{4} b_4) \tilde{\nabla}^{a}\sigma \tilde{\nabla}_{b}\tilde{\nabla}_{a}\sigma \tilde{\nabla}^{b}\sigma  \nonumber \\ 
&+ \tfrac{1}{16} (64 b_1 + 32 b_2 + 48 b_3 - 4 b_4 - 4 b_5 - 8 b_6  ) \tilde{\nabla}_{a}\sigma \tilde{\nabla}^{a}\sigma \tilde{\nabla}_{b}\sigma \tilde{\nabla}^{b}\sigma  \nonumber \\ 
&-  b_4 \tilde{\nabla}^{a}P \tilde{\nabla}_{b}\tilde{\nabla}_{a}\sigma \tilde{\nabla}^{b}\sigma - (3b_8+8b_7-16b_6-8b_5+32b_3) \tilde{\nabla}_{a}P \tilde{\nabla}^{a}P \tilde{\nabla}_{b}\sigma \tilde{\nabla}^{b}P  \nonumber \\ 
& + \tfrac{1}{2} (-8 b_3 + b_6) \tilde{R} \tilde{\nabla}_{a}\tilde{\nabla}^{a}\sigma+ (4 b_1+ b_2) \tilde{\nabla}_{b}\tilde{\nabla}_{a}\sigma \tilde{\nabla}^{b}\tilde{\nabla}^{a}\sigma
 \Big)\labell{eq.2.35}
\end{align}
 to be invariant under the T-duality \cite{Razaghian:2017okr}. That is, when applying these corrections on the leading order $d$-dimensional couplings in $\bS_0$, 
the resulting field redefinitions terms guarantee that the couplings at order $\alpha'$ in $\bS_1$   are   invariant under the Buscher rules, \ie
\beqa
&&\bS_0\Big(-\sigma+\alpha'\delta\sigma^{(1)},P+\alpha'\delta P^{(1)},g +\alpha'\delta g^{(1)} \Big)-\bS_0(\sigma,P,g )\nn\\
&&+\bS_1(-\sigma,P,g )-\bS_1(\sigma,P,g )=0\labell{ds1}
\eeqa
In the perturbation of the first term, one must ignore the terms at order $\alpha'^2$ and higher. The corrections to the Buscher rules at order $\alpha'$, \ie \reef{eq.2.2.20}, should satisfy the above constraint. In solving this  constraint, one must add all total derivative terms at order $\alpha'$ to the above constraint. The result is \cite{Razaghian:2017okr}
\begin{align}
& A_{1} =  \tfrac{1}{8} ( 4 A_6  - A_{9}(D-3) + 2 b_4 + 4 b_5 + 4 b_6 ), \nonumber\\
&   A_{2} = \tfrac{1}{2} \bigl(4 A_6    -  A_{9} ( D-2)- 8 b_2 + 3 b_4 + 2 b_7 + b_8\bigr), \nonumber\\
& A_3=\tfrac{1}{2} \bigl(-4 A_6  + A_{9} ( D-1) + 16 b_2 - 32 b_3 - 5 b_4 + 8 b_5 + 16 b_6 - 8 b_7 - 3 b_8 \bigr),\nonumber\\
& A_{4}= \tfrac{1}{8} \bigl(-4 A_6  + A_{9} ( D-3) + 32 b_1 - 32 b_3 - 3 b_4 + 8 b_6 - 4 b_7 -  b_8 \bigr), \nonumber\\
& A_{5} =  \tfrac{1}{8} (-4 A_6   + 8 b_2 +32(D-2) b_3 +(D- 5) b_4 -4(D-2) b_5 - 4(3D-7) b_6 \nn\\
&\qquad+4(D- 3) b_7 +(D- 3) b_8 ), \nonumber\\
&  A_{7} =  8 b_2 - 2 b_4, \nonumber\\
& A_{8}=\tfrac{1}{2} (- A_{9} + 32 b_3 + b_4 - 4 b_5 - 12 b_6 + 4 b_7 + b_8) \labell{c1}
\end{align}
 The residual parameters $A_6,A_9$ are not fixed by the calculations at order $\alpha'$ and $\alpha'^2$  that have been done in \cite{Razaghian:2017okr}. In the Heterotic theory, we will see that these parameters as well as the parameters $b_4,b_5,b_8$ will be fixed by requiring the couplings at order  $\alpha'^2$ which are produced by the Green-Schwarz mechanism \cite{Green:1984sg}, to be invariant under the T-duality at order $\alpha'$.
 
Applying  the    variations \reef{c1}   to the couplings $\bS_1$, one finds some couplings at order $\alpha'^2$.  On the other hand,  it is known that there is no curvature couplings at order $\alpha'^2$ in the Heterotic theory, hence,  there must be corrections to the Buscher rules at order $\alpha'^2$ as well. The effect of applying these corrections to the couplings $\bS_0$ must be canceled by the effect of applying the corrections at order $\alpha'$ on the couplings in $\bS_1$. Therefore, the corrections to the Buscher rules at orders $\alpha'$ and $\alpha'^2$, \ie
\beqa
\sigma&\rightarrow &-\sigma+\alpha'\delta\sigma^{(1)}+\alpha'^2\delta\sigma^{(2)}\nn\\
P&\rightarrow &P+\alpha'\delta P^{(1)}+\alpha'^2\delta P^{(2)}\nn\\
g_{ab}&\rightarrow &g_{ab}+\alpha'\delta g^{(1)}_{ab}+\alpha'^2\delta g^{(2)}_{ab}\labell{per12}
\eeqa
must satisfy the   following  constraint: 
\beqa
&&\bS_0\Big(-\sigma+\alpha'\delta\sigma^{(1)}+\alpha'^2\delta\sigma^{(2)},P+\alpha'\delta P^{(1)}+\alpha'^2\delta P^{(2)},g +\alpha'\delta g^{(1)} +\alpha'^2\delta g^{(2)}  \Big)\nn\\
&&-\bS_0(\sigma,P,g )+\bS_1(-\sigma+\alpha'\delta\sigma^{(1)},P+\alpha'\delta P^{(1)},g +\alpha'\delta g^{(1)}  )-\bS_1(\sigma,P,g )=0\labell{ds2}
\eeqa
In the perturbation of the first and the third terms, one must ignore the terms at order $\alpha'^3$ and higher. In solving the above constraint, one must add to it all total derivative terms at order $\alpha'^2$. Using the fact that the T-duality transformations must be a $ \mathbb{Z}_2 $-group, one finds that there are 98 coefficients in  the variations  $\delta \sigma^{(2)} , \delta P^{(2)} ,  \delta g^{(2)} _{ab}$. The above constraint fix 61 coefficients in terms of   other 37 terms and in terms of the $b$-coefficients \cite{Razaghian:2017okr}.

In order to study the couplings at order $\alpha'^3$ under the T-duality, one must consider corrections to the Buscher  rules at order $\alpha'^3$ as well, \ie
\beqa
\sigma&\rightarrow &-\sigma+\alpha'\delta\sigma^{(1)}+\alpha'^2\delta\sigma^{(2)}+\alpha'^3\delta\sigma^{(3)}\nn\\
P&\rightarrow &P+\alpha'\delta P^{(1)}+\alpha'^2\delta P^{(2)}+\alpha'^3\delta P^{(3)}\nn\\
g_{ab}&\rightarrow &g_{ab}+\alpha'\delta g^{(1)}_{ab}+\alpha'^2\delta g^{(2)}_{ab}+\alpha'^3\delta g^{(3)}_{ab}\labell{per123}
\eeqa
A straightforward extension of  the constraint \reef{ds2}  to order $\alpha'^3$ is given by the following constraint:
\beqa
&&\bS_0\Big(-\sigma+\alpha'\delta\sigma^{(1)}+\alpha'^2\delta\sigma^{(2)}+\alpha'^3\delta\sigma^{(3)},P+\alpha'\delta P^{(1)}+\alpha'^2\delta P^{(2)}+\alpha'^3\delta P^{(3)}\nn\\
&&,g +\alpha'\delta g^{(1)} +\alpha'^2\delta g^{(2)} +\alpha'^3\delta g^{(3)} \Big)-\bS_0(\sigma,P,g )\nn\\
&&+\bS_1\Big(-\sigma+\alpha'\delta\sigma^{(1)}+\alpha'^2\delta\sigma^{(2)},P+\alpha'\delta P^{(1)}+\alpha'^2\delta P^{(2)},g +\alpha'\delta g^{(1)} +\alpha'^2\delta g^{(2)} \Big)\nn\\
&&-\bS_1(\sigma,P,g )+\bS_3(-\sigma,P,g)-\bS_3(\sigma,P,g)=0\labell{cons3}
\eeqa
where $\bS_3(-\sigma,P,g)-\bS_3(\sigma,P,g)=\delta\bS_3$ is \reef{sons}.  In the perturbation of the first and the third terms, one must ignore the terms at order $\alpha'^4$ and higher. The coefficients of the variations $\delta \sigma^{(1)} , \delta P^{(1)} ,  \delta g^{(1)} _{ab}$ are given in \reef{c1} and of the variations $\delta \sigma^{(2)} , \delta P^{(2)} ,  \delta g^{(2)} _{ab}$ satisfy the constraint   \reef{ds2}. After solving the constraint \reef{ds2}, one must replace the corresponding variations into the above constraint.

Adding   all total derivative terms  to the constraint \reef{cons3}, and writing them in terms of independent couplings, one finds many algebraic equations involving the $d$-coefficients, the $b$-coefficients, the coefficients of the total derivative terms, and the coefficients of the variations. Solving these equations, one finds 14 relations between the 37 unfixed coefficients of  $\delta \sigma^{(2)} , \delta P^{(2)} ,  \delta g^{(2)} _{ab}$. Moreover, one  finds   many coefficients in the  variations  $\delta \sigma^{(3)} ,\ \delta P^{(3)} ,  \delta g^{(3)} _{ab}$ are not fixed, and the remaining  coefficients are fixed in terms of the unfixed coefficients, the $d$-coefficients and the $b$-coefficients.    However,  the equations fix   the $d$-coefficients in the effective action \reef{RRRR} in terms of the $b$-coefficients, \ie
\beqa
 d_1+2d_4&\!\!\!\!\!=\!\!\!\!\!&-\frac{d_2}{4}+\frac{1}{256}(8b_3-b_5-2b_6)(16b_1+8b_3-b_5-2b_6)(8b_1+28b_2+108b_3-18b_6+3b_7)\nn\\
d_5&\!\!\!\!\!=\!\!\!\!\!&-\frac{b_1}{4}(2b_1+b_2)(16b_1+8b_3-b_5-2b_6)\nn\\
d_7&\!\!\!\!\!=\!\!\!\!\!&-\frac{b_1}{64}(16b_1+8b_3-b_5-2b_6)(8b_2+36b_3-6b_6+b_7)\,\,;\,\,\,d_3=d_6=0\labell{d5d7}
\eeqa
The only unknown $d$-coefficient at order $\alpha'^3$  is $d_2$. 

As we have anticipated before, the coefficients $d_5,d_7$ which are changed under field redefinitions in the Heterotic theory, depend on the form of effective action at order $\alpha'$. However, the Riemann curvature couplings in \reef{RRRR} with coefficients $d_1,d_4,d_2$ are not changed under field redefinitions, hence, we do not expect these coefficients to depend on the effective action \reef{finalS10}. Therefore, we expect the coefficients $b_2,b_3,\cdots, b_8$ in \reef{finalS10} not to be totally arbitrary. The invariance of the curvature terms at order $\alpha'^2,\alpha'^3$ under T-duality does not constraint these coefficients. However, the Heterotic theory has another gravity couplings which is resulted from the Green-Schwarz mechanism \cite{Green:1984sg}. These couplings may constrain the parameters $b_2,b_3,\cdots, b_8$.

Extension of the effective action at the leading order of $\alpha'$, \ie \reef{eq.1.110},    in the presence of B-field is
\begin{align}
S_0=-\frac{2}{\kappa^2}\int d^{d+1}x e^{-2\Phi}\sqrt{-G}\,  \left(R + 4\nabla_{\alpha}\Phi \nabla^{\alpha}\Phi-\frac{1}{12}H^2\right)\,.\labell{S0b}
\end{align}
where $H=dB$. This action has been written in DFT formalism in \cite{Hull:2009mi}. In the Heterotic theory, the Green-Schwarz mechanism \cite{Green:1984sg} dictates that the B-field 
  strength $ H(B) $ must be replaced by the improved field
strength $ \widehat{H}(B,\Gamma) $ that  includes  the  Chern-Simons term  built  from  the  Christoffel  connection:
\begin{align}
\widehat{H}_{\mu\nu\rho}(B,\Gamma)=3(\partial_{[\mu}B_{\nu\rho]}+\alpha'\Omega(\Gamma)_{\mu\nu\rho})
\end{align}
with the Chern-Simons three-form
\begin{align}
\Omega(\Gamma)_{\mu\nu\rho}=\Gamma^\alpha_{[\mu|\beta|}\partial_\nu\Gamma^\beta_{\rho]\alpha}+\frac{2}{3}\Gamma^\alpha_{[\mu|\beta|}\Gamma^\beta_{\nu|\gamma|}\Gamma^\gamma_{\rho]\alpha}\,.
\end{align}
The replacement $H\rightarrow \widehat{H}$ in $S_0$ produces the gravity coupling  
$\alpha'^2 \Omega^2$ which should be invariant under T-duality.
 
The effective action at order $\alpha'$, \ie \reef{finalS10},  in the presence of B-field  is \cite{Metsaev:1987zx,Bergshoeff:1989de}
\beqa
 S_1&=&\frac{-2b_1}{\kappa^2}\alpha'\int d^{d+1}x e^{-2\Phi}\sqrt{-G}\Big(   R_{\alpha \beta \gamma \delta} R^{\alpha \beta \gamma \delta} -\frac{1}{2} R_{\alpha \beta \gamma \delta} H^{\alpha \beta\lambda}H^{ \gamma \delta}{}_{\lambda}\nn\\
&&+\frac{1}{24}H_{\mu\nu\rho}H^{\mu}{}_{\alpha}{}^{\beta}H^{\nu}{}_{\beta}{}^{\gamma}H^{\rho}{}_{\gamma}{}^{\alpha}-\frac{1}{8}H_{\mu}{}^{\alpha\beta}H_{\nu\alpha\beta}H^{\mu\gamma\rho}H^{\nu}{}_{\gamma\rho}+\cdots\Big)\labell{S1b}
\eeqa
where dots represent the terms which can be  removed  by appropriate field redefinitions. The DFT formulation of this  action   has been   found in \cite{Marques:2015vua,Baron:2017dvb}.  The $H$ in the Heterotic theory must be replaced by $\widehat{H}$. This replacement  produces the gravity couplings $\alpha'^3 R_{\alpha \beta \gamma \delta} \Omega^{\alpha \beta\lambda}\Omega^{ \gamma \delta}{}_{\lambda}$ and also some $ \Omega^4$ terms in which we are not interested in this paper because they are at order $\alpha'^5$. The consistency of our calculations require these gravity couplings to be invariant under the T-duality transformations too.

Reduction of   $ \Omega^2$ from 10-dimensional to 9-dimensional spacetime is
\beqa
 \Omega^2&=&- \tfrac{8}{9} \Gamma ^{abc} \Gamma _{b}{}^{de} \Gamma _{da}{}^{i} \Gamma ^{j}{}_{c}{}^{k} \Gamma _{ki}{}^{l} \Gamma _{lej} + \tfrac{8}{9} \Gamma ^{abc} \Gamma _{b}{}^{de} \Gamma _{da}{}^{i} \Gamma ^{j}{}_{c}{}^{k} \Gamma _{ke}{}^{l} \Gamma _{lij} -  \tfrac{1}{6} \Gamma ^{abc} \Gamma ^{dei} \tilde{\nabla}_{c}\Gamma _{idj} \tilde{\nabla}_{e}\Gamma _{ba}{}^{j}\nn\\&& -  \tfrac{4}{3} \Gamma ^{abc} \Gamma _{b}{}^{de} \Gamma _{da}{}^{i} \Gamma ^{j}{}_{c}{}^{k} \tilde{\nabla}_{e}\Gamma _{kij} + \tfrac{4}{3} \Gamma ^{abc} \Gamma _{b}{}^{de} \Gamma _{da}{}^{i} \Gamma ^{j}{}_{c}{}^{k} \tilde{\nabla}_{i}\Gamma _{kej} + \tfrac{1}{3} \Gamma ^{abc} \Gamma ^{dei} \tilde{\nabla}_{c}\Gamma _{idj} \tilde{\nabla}^{j}\Gamma _{bae}\nn\\&& -  \tfrac{1}{6} \Gamma ^{abc} \Gamma ^{dei} \tilde{\nabla}_{j}\Gamma _{icd} \tilde{\nabla}^{j}\Gamma _{bae} -  \tfrac{1}{6} \Gamma ^{abc} \Gamma ^{d}{}_{b}{}^{e} \tilde{\nabla}_{i}\Gamma _{edj} \tilde{\nabla}^{j}\Gamma _{ca}{}^{i} + \tfrac{1}{6} \Gamma ^{abc} \Gamma ^{d}{}_{b}{}^{e} \tilde{\nabla}_{j}\Gamma _{edi} \tilde{\nabla}^{j}\Gamma _{ca}{}^{i}\labell{Oco}
\eeqa
As can be seen, it contains no term which has $ \sigma $. So it is invariant under the Buscher rules \reef{buscher}. It must be  also invariant under the T-duality transformations at order $\alpha'$, \ie  
\beqa
e^{-2 P}\sqrt{-g}\Omega^2(P+\alpha' \delta P^{(1)},g+\alpha'\delta g^{(1)})-e^{-2 P}\sqrt{-g}\Omega^2(P,g )&=&0\labell{Ocos}
\eeqa
 This constraint fixes the residual parameters in \reef{c1} to be zero, \ie $A_6=A_9=0$, and also fix the coefficients $b_4,b_5,b_8$ in terms of $b_2,b_3,b_6,b_7$, \ie
\beqa
b_4=4b_2\,\,\,;\,\,\, b_5=8b_3-2b_6\,\,\,;\,\,\,b_8=-4(b_2-b_6+b_7)\labell{3b}
\eeqa
 Hence the corrections to the Buscher rules at order $\alpha'$, \ie \reef{c1},  is fixed to be
\beqa
\delta\sigma^{(1)} &=&(b_2+3b_3-\frac{1}{2}b_6)R+(2b_6-b_7)\tilde{\nabla}_a\tilde{\nabla}^a P+(4b_1-b_2-4b_3+\frac{1}{2}b_6 )\tilde{\nabla}_{a}\sigma \tilde{\nabla}^{a}\sigma \nonumber\\
\delta P^{(1)} &=&0\nonumber\\
\delta g_{ab}^{(1)}&=& 0 \labell{00}
\eeqa
Note that the $d$-dimensional couplings in $\Omega^2$ can not be written in terms of Riemann curvatures, hence, the constraint \reef{Ocos} is independent of the constraint \reef{cons3}. 

Let us compare the above transformation with the standard T-duality transformation at order $\alpha'$ \cite{Bergshoeff:1995cg} when the effective action has no Ricci or scalar  curvature. The constraint that the effective action at order $\alpha'$ in the Heterotic theory must be invariant under the T-duality, has been used in \cite{Bergshoeff:1995cg} to find the   extension of the Buscher rules at order $\alpha'$ in the presence of B-field and gauge field.   In the absence of these fields, and for diagonal metric, they are \cite{Bergshoeff:1995cg}  
\begin{align}
\tilde{g}_{ab}&=g_{ab}+\frac{g_{yy}\hat{\mathcal{G}}_{ya}\hat{\mathcal{G}}_{yb}-2\hat{\mathcal{G}}_{yy}\hat{\mathcal{G}}_{y(a}g_{b)y}}{\hat{\mathcal{G}}^2_{yy}}\\
\tilde{\Phi}&=\Phi-\frac{1}{2}\log|\hat{\mathcal{G}}_{yy}|\nn\\
\tilde{g}_{yy}&=e^{2\tilde{\sigma}} =\frac{e^{2\sigma}}{\hat{\mathcal{G}}^2_{yy}}\nn
\end{align}
where the $\alpha'$-correction appears in $\hat{\mathcal{G}}_{\mu\nu}$, \ie
\begin{align}
\hat{\mathcal{G}}_{\mu\nu}=G_{\mu\nu}+\frac{1}{4}\alpha' \Omega_\mu{}^{\bar{a}\bar{b}} \Omega_\nu{}^{\bar{a}\bar{b}} 
\end{align}
The metric $G_{\mu\nu}$ is the 10-dimensional metric and $\omega_\mu{}^{\bar{a} \bar{b}}$ is torsionless spin connection, \ie 
\begin{align}
 \Omega_\mu{}^{\bar{a} \bar{b}} &=\omega_\mu{}^{\bar{a} \bar{b}} =e_\alpha{}^{\bar{a}} e^{\lambda \bar{b}}\Gamma^\alpha_{\mu\lambda}-e^{\lambda \bar{b}}\partial_\mu e_\lambda{}^{\bar{a}}\nn
\end{align}
Using the fact that fields are independent of the $y$-direction, one finds that $\Omega_a{}^{\bar{a}\bar{b}}  \Omega_y{}^{\bar{a}\bar{b}}\,=0=\, \hat{\mathcal{G}}_{ay}$. One  also finds $\hat{\mathcal{G}}_{yy}=e^{2\sigma}\Big(1-\frac{1}{2}\alpha'\tilde{\nabla}_a\sigma \tilde{\nabla}^a\sigma\Big)$. Hence, the 9-diemsional metric and dilaton become invariant and  the transformation of $\sigma$ becomes    the same as the transformation \reef{00} in which $b_2=b_3=b_6=b_7=0$.

The $R\Omega^2$ couplings are at order $\alpha'^3$, so to the order that we consider in this paper, the consistency requires it to be invariant under the Buscher rules. The reduction of this term to the $d$-dimensional spacetime is  
\beqa
 R_{\alpha \beta \gamma \delta} \Omega^{\alpha \beta\lambda}\Omega^{ \gamma \delta}{}_{\lambda}&=&\tfrac{16}{9} \Gamma ^{abc} \Gamma _{b}{}^{de} \Gamma _{da}{}^{i} \Gamma ^{j}{}_{c}{}^{k} \Gamma _{k}{}^{lm} \Gamma _{lj}{}^{n} \tilde{R}_{eimn} + \tfrac{8}{9} \Gamma ^{abc} \Gamma _{b}{}^{de} \Gamma _{da}{}^{i} \Gamma ^{jkl} \tilde{R}_{eilm} \tilde{\nabla}_{c}\Gamma _{kj}{}^{m} \nn\\&&+ \tfrac{1}{9} \Gamma ^{abc} \Gamma ^{dei} \tilde{R}_{cjil} \tilde{\nabla}_{k}\Gamma _{ed}{}^{l} \tilde{\nabla}^{k}\Gamma _{ba}{}^{j} + \tfrac{1}{9} \Gamma ^{abc} \Gamma ^{dei} \tilde{R}_{ckil} \tilde{\nabla}^{k}\Gamma _{ba}{}^{j} \tilde{\nabla}^{l}\Gamma _{edj}\nn\\&& -  \tfrac{2}{9} \Gamma ^{abc} \Gamma ^{dei} \tilde{R}_{cjil} \tilde{\nabla}^{k}\Gamma _{ba}{}^{j} \tilde{\nabla}^{l}\Gamma _{edk} + \tfrac{1}{9} \Gamma ^{abc} \Gamma ^{d}{}_{b}{}^{e} \tilde{R}_{ijkl} \tilde{\nabla}^{j}\Gamma _{ca}{}^{i} \tilde{\nabla}^{l}\Gamma _{ed}{}^{k} \nn\\&&+ \tfrac{2}{9} \Gamma ^{abc} \Gamma ^{dei} \tilde{R}_{cjkl} \tilde{\nabla}_{e}\Gamma _{ba}{}^{j} \tilde{\nabla}^{l}\Gamma _{id}{}^{k} -  \tfrac{2}{9} \Gamma ^{abc} \Gamma ^{dei} \tilde{R}_{cjkl} \tilde{\nabla}^{j}\Gamma _{bae} \tilde{\nabla}^{l}\Gamma _{id}{}^{k} \nn\\&&-  \tfrac{8}{9} \Gamma ^{abc} \Gamma _{b}{}^{de} \Gamma _{da}{}^{i} \Gamma ^{jkl} \tilde{R}_{eilm} \tilde{\nabla}^{m}\Gamma _{kcj} + \tfrac{8}{9} \Gamma ^{abc} \Gamma _{b}{}^{de} \Gamma _{da}{}^{i} \Gamma ^{j}{}_{c}{}^{k} \tilde{R}_{eilm} \tilde{\nabla}^{m}\Gamma _{kj}{}^{l}\labell{ROO}
\eeqa
Since $\sigma$ does not appear in it, it is obviously invariant under the Buscher rules \reef{buscher}.

The constraints \reef{3b}, simplify the equations in \reef{d5d7} as
\beqa
 d_1+2d_4&=&-\frac{d_2}{4} \nn\\
d_5&=&-4b_1^2(2b_1+b_2) \nn\\
d_7&=&-\frac{b_1^2}{4} (8b_2+36b_3-6b_6+b_7)\,\,;\,\,\,d_3=d_6=0\labell{d5d71}
\eeqa
As expected, the $b$-coefficients do not appear in the first equation. Moreover, for the specific field variables at order $\alpha'$, \ie $b_2=-2b_1,\, 36b_3-6b_6+b_7=16b_1$, the Riemann curvature couplings with coefficients $d_5,d_7$ should be removed by the field redefinitions. 

  Since, our calculations in the absence of B-field can not fix the coefficient $d_4$, we have to fix it by hand. In the superstring theory, we showed that  $d_4=0$   precisely reproduces the known $R^4$ corrections to the type II supergravity. The difference between the superstring and the Heterotic calculations is the presence of effective action   at order $\alpha'$. The presence of this action   may cause the coefficient  $d_4$ not to be zero in the Heterotic theory. If we choose it to be $d_4=-2b_1^3$, then the equations \reef{d5d71} produce  the couplings \reef{II} as well as the following couplings:
\beqa
S^H_3&\!\!\!\!\!=\!\!\!\!\!&-\frac{2b_1^2}{\kappa^2}\int d^{d+1}x e^{-2\Phi}\sqrt{-G}\bigg[ 4b_1 R_{\alpha \beta}{}^{\zeta \eta} R^{\alpha \beta \gamma \delta} R_{\gamma \zeta}{}^{\theta \iota} R_{\delta \eta \theta \iota}-2b_1 R_{\alpha \beta}{}^{\zeta \eta} R^{\alpha \beta \gamma \delta} R_{\gamma \delta}{}^{\theta \iota} R_{\zeta \eta \theta \iota} \labell{RRRR21}\\
&&  -4(2b_1+b_2) R_{\alpha \beta \gamma}{}^{\zeta} R^{\alpha \beta \gamma \delta} R_{\delta}{}^{\eta \theta \iota} R_{\zeta \eta \theta \iota}   -\frac{1}{4} (8b_2+36b_3-6b_6+b_7)(R_{\mu\nu\alpha\beta}R^{\mu\nu\alpha\beta})^2  \bigg]\nn
\eeqa
Since the last two terms above are changed under field redefinitions, we have to choose a specific field variable to compare them with the couplings in the literature. To compare the couplings with the couplings that have been found by the S-matrix method, one has to choose the effective action at order $\alpha'$ in a specific field variables that do not change the graviton and dilaton propagators.  That is, we have to choose the Gauss-Bonnet combinations for the curvature couplings at order $\alpha'$, \ie $b_2=-4b_1,\,b_3=b_1$, to have standard graviton propagator, and also we have to choose $b_6=b_7=0$ to have standard dilaton propagator.  For these parameters, the above couplings becomes
\beqa
S^H_3&\!\!\!\!\!=\!\!\!\!\!&-\frac{2b_1^3}{\kappa^2}\int d^{d+1}x e^{-2\Phi}\sqrt{-G}\bigg[ 4  R_{\alpha \beta}{}^{\zeta \eta} R^{\alpha \beta \gamma \delta} R_{\gamma \zeta}{}^{\theta \iota} R_{\delta \eta \theta \iota}-2  R_{\alpha \beta}{}^{\zeta \eta} R^{\alpha \beta \gamma \delta} R_{\gamma \delta}{}^{\theta \iota} R_{\zeta \eta \theta \iota} \labell{RRRR2}\\
&& + 8 R_{\alpha \beta \gamma}{}^{\zeta} R^{\alpha \beta \gamma \delta} R_{\delta}{}^{\eta \theta \iota} R_{\zeta \eta \theta \iota}   - (R_{\mu\nu\alpha\beta}R^{\mu\nu\alpha\beta})^2  \bigg]\nn
\eeqa
  Using the tensor \reef{t8}, one can write  
\beqa
t^{\mu_1\cdots\mu_8}\Tr(R_{\mu_1\mu_2}R_{\mu_3\mu_4})\Tr(R_{\mu_5\mu_6}R_{\mu_7\mu_8})&=&8 R_{\alpha \beta}{}^{\zeta \eta} R^{\alpha \beta \gamma \delta} R_{\gamma \zeta}{}^{\theta \iota} R_{\delta \eta \theta \iota}   -4 R_{\alpha \beta}{}^{\zeta \eta} R^{\alpha \beta \gamma \delta} R_{\gamma \delta}{}^{\theta \iota} R_{\zeta \eta \theta \iota}\nn\\
&&  +16 R_{\alpha \beta \gamma}{}^{\zeta} R^{\alpha \beta \gamma \delta} R_{\delta}{}^{\eta \theta \iota} R_{\zeta \eta \theta \iota}-2(R_{\mu\nu\alpha\beta}R^{\mu\nu\alpha\beta})^2 \nn
\eeqa
Therefor, the effective actions that the T-duality constraint produces in the Heterotic theory for the specific parameters $b_2=-4b_1,\,b_3=b_1,\,b_6=b_7=0$ are \reef{Y1} and 
\beqa
S^H_3&=&-\frac{b_1^3}{\kappa^2}\int d^{d+1}x e^{-2\Phi}\sqrt{-G}\bigg[ t^{\mu_1\cdots\mu_8}\Tr(R_{\mu_1\mu_2}R_{\mu_3\mu_4})\Tr(R_{\mu_5\mu_6}R_{\mu_7\mu_8})  \bigg]\labell{H2}
\eeqa
Which are exactly   the couplings that have been found in  \cite{Gross:1986mw}. 

We have seen that the gravity couplings resulted from the Green-Schwarz mechanism fix the residual T-duality parameters at order $\alpha'$ and also fixes the parameters $b_4,b_5,b_8$. There are also 23 T-duality parameters at order $\alpha'^2$ that are not fixed by the constraint \reef{cons3}. These parameter may also be fixed by the  gravity couplings resulting from the Green-Schwarz mechanism. Since these mechanism does not produce  gravity couplings at order $\alpha'^4$, one expects the T-duality  transformation of $\alpha'^2\Omega^2$ at order $\alpha'^2$ cancels the T-duality transformation of $\alpha'^3R\Omega^2$ at order $\alpha'$. On the other hand,  there is no $\sigma$ in the reduction of $\alpha'^3R\Omega^2$, \ie \reef{ROO},  and the T-duality transformation \reef{00} at order $\alpha'$ does not change $P$ and $g_{ab}$, so  $\alpha'^3R\Omega^2$ is invariant under the T-duality transformation at order $\alpha'$. Therefore, the T-duality transformation of $\alpha'^2\Omega^2$ at order $\alpha'^2$ must be zero,\ie
\beqa
e^{-2 P}\sqrt{-g}\Omega^2(P+\alpha' \delta P^{(1)}+\alpha'^2 \delta P^{(2)},g+\alpha'\delta g^{(1)}+\alpha'^2\delta g^{(2)})-e^{-2 P}\sqrt{-g}\Omega^2(P,g )&=&0 
\eeqa
 This may further fix the parameters $b_2,b_3,b_6,b_7$ in the effective action at order $\alpha'$ and the residual T-duality parameters at order $\alpha'^2$. It would be interesting to perform this calculations in details. It would be also interesting to extend the calculations in this paper which have no B-field,   to the case that the B-field is non-zero. That calculation would produce the B-field couplings at order $\alpha'^3$, \ie the extension of \reef{S1b} to order $\alpha'^3$, which is not known in the literature.
  The  T-duality transformations at order $\alpha'$ in the presence of B-field have  been found in \cite{Bergshoeff:1995cg,Kaloper:1997ux}.

{\bf Acknowledgments}:   This work is supported by Ferdowsi University of Mashhad under grant  2/45608(1396/10/05).

\end{document}